\documentclass{webofc}
\usepackage{bm}
\usepackage{bbm}
\usepackage{hyperref}

\def\la{\lambda} \def\lap{\lambda^{\prime}} \def\pa{\partial} \def\de{\delta}  \def\dag{\dagger}
    \def\S{{\rm S}}
\def\bnabla{{\bm \nabla}}

\def\nn{\nonumber}

\begin{document}
\title{Heavy hybrids and tetraquarks in effective field theory}

\author{\firstname{Jaume} \lastname{Tarr\'us Castell\`a}\inst{1}\fnsep\thanks{\email{jtarrus@ifae.es}}}

\institute{Grup de F\'\i sica Te\`orica, Dept. F\'\i sica and IFAE-BIST, 
Universitat Aut\`onoma de Barcelona, \\ E-08193 Bellaterra (Barcelona), Spain}

\abstract{
We report on an effective field theory (EFT) description of exotic quarkonia as bound states on the spectrum of hybrid and tetraquark static energies. We provide expressions for hybrid and tetraquark static energies in terms of Wilson loops. The former have been computed in quenched lattice calculations but the latter are yet unavailable. From the few simulations with dynamical light-quarks we argue that the overall picture from hybrid static energies does not change but additional states, such as heavy meson pairs, need to be considered for a full description. In this EFT framework for quarkonium hybrids, we report on recent results for mixing with standard quarkonium, spin-dependent contributions, and semi-inclusive decays.
}
\maketitle
\section{Introduction} \label{intro}
During the past 15 years, experimental observations have revealed the existence of a large number of unexpected states in the quarkonium spectrum that do not fit standard quarkonium description. These exotic states have attracted broad experimental and theoretical attention because they are candidates for nontraditional hadronic states, that is states that cannot be classified as mesons or baryons. Many phenomenological models for exotics have been proposed which can be loosely classified into two categories, those containing an excited gluonic state and the ones containing four constituent quarks. The latter can be further subdivided corresponding to different spatial arrangements of the four constituent quarks. However a compelling and unified understanding of these new states has not yet emerged. 

EFTs exploit the separation of scales in a system. In the case of exotic quarkonium we have at least two large energy gaps~\cite{Brambilla:2017uyf, Oncala:2017hop}. First, the heavy-quarks are nonrelativistic and $m\gg\Lambda_{\rm QCD}$, where $m$ is the heavy quark mass. Therefore exotic quarkonium can be studied in NRQCD. In the static limit, the spectrum of the theory is formed by the so-called static energies. In the quenched approximation, this spectrum is the well-known hybrid static energy spectrum~\cite{Juge:1997nc,Juge:2002br,Bali:2000vr,Capitani:2018rox}. In the case with dynamic light-quarks new states appear, most prominently heavy-meson pairs~\cite{Bali:2003jq} and we expect the appearance of tetraquark static energies \cite{Brambilla:2008zz,Braaten:2014qka} as well as single-heavy-baryon pairs and quarkonium and light-quark baryon states. The exotic states appear as heavy-quark antiquark bound states around the minima of these static energies. This leads naturally to an energy gap between the heavy-quark antiquark binding energies $E_b$ and the light-quark and gluon dynamical energy scale $\Lambda_{\rm QCD}\gg E_b$. This scale separation has lead to the observation that exotic quarkonia can be studied in a Born-Oppenheimer picture. The static energies, as well as the other matching coefficients of the low-energy EFT describing the heavy-quark bound states, can be written in terms of static Wilson loops which can be computed on the lattice. In the short-distance regime $r\lesssim\Lambda_{\rm QCD}$ the relative momentum of the heavy quarks can also be integrated out perturbatively leading to a short-distance description of the matching coefficients. In the long-distance regime $r\gg 1/\Lambda_{\rm QCD}$ Effective String Theories can be used to model the long-distance part of the potentials~\cite{Nambu:1978bd,Polchinski:1991ax,Luscher:2002qv,PerezNadal:2008vm,Oncala:2017hop}.

In these proceedings we briefly outline the construction of EFTs for quarkonium hybrids and tetraquarks and sumarize some of the most recent results.

\section{The effective field theory}\label{sec-1}

Exotic quarkonia are characterized by being formed by two distinct components: on one hand we have the heavy-quark-antiquark pair and on the other gluonic or light-quark degrees of freedom. The heavy-quark-antiquark pair forms a nonrelativistic bound state with three characteristic scales, $m$ the heavy quark mass, $mv$ the relative momentum, with $v\ll 1$ the relative velocity, and $mv^2$ the heavy quark binding energy. These scales fulfill the hierarchy $m\gg mv \gg mv^2$. The light degrees of freedom are characterized by a typical energy and momentum of order $\Lambda_{\rm QCD}$.
This implies that the typical size of exotic quarkonia is of order $1/\Lambda_{\rm QCD}$. The scaling of the typical distance of the heavy quark-antiquark pair $r\sim 1/(mv)$ depends on the details of the full inter-quark potential, which has a long-range nonperturbative part and a short-range Coulomb interaction. In the most general assumption is $r\lesssim 1/\Lambda_{\rm QCD}$ (or $mv\gtrsim \Lambda_{\rm QCD}$). We can use the separation of the scales on the system to build EFTs to describe Exotic quarkonia~\cite{Berwein:2015vca, Oncala:2017hop, Brambilla:2017uyf}.
 
NRQCD~\cite{Caswell:1985ui,Bodwin:1994jh} is obtained by integrating it out the heavy quark mass, $m$. We can study the spectrum of states with a heavy-quark and antiquark at leading order in NRQCD (the static limit). In the static limit the eigenstates are characterized by quark-antiquark separation, the flavor content of the light degrees of freedom (for simplicity we will just consider isospin) and the quantum numbers corresponding to representations of $D_{\infty h}$. According to this symmetry, the mass eigenstates are classified in terms of the angular momentum along the quark-antiquark axis ($\Lambda=0,\,1,\,2\dots$, to which one gives the traditional names $\Sigma$, $\Pi$, $\Delta$), $CP$ ($g$ for even or $u$ for odd), and the reflection properties with respect to a plane that passes through the quark-antiquark axis (+ for even or - for odd). Only the $\Sigma$ states are not degenerate with respect to the reflection symmetry.

The specific form of these static eigenstates depends on nonperturbative physics and are unknown, nevertheless the corresponding energy eigenvalues, the static energies, can be obtained from large time logarithms of appropriate correlators
\begin{align}
E^{(0)}_{n}(r)=\lim_{T\to\infty}\frac{i}{T}\log \langle \mathcal{O}_n(T,\,\bm{r},\,\bm{R})|\mathcal{O}_n(0,\,\bm{r},\,\bm{R})\rangle\,,\label{ste}
\end{align}
where $n$ stands for the set of quantum numbers that identify the static eigenstate, $\bm{R}$ and $\bm{r}$ are the center of mass and relative coordinates of the heavy-quark pair and $\mathcal{O}_n$ is an interpolating operator.

For hybrid and tetraquark states an appropriate interpolating operator reads
\begin{align}
\mathcal{O}_n(t,\,\bm{r},\,\bm{R})=\chi(t,\,\bm{R}-\bm{r}/2)\phi(t,\,\bm{R}-\bm{r}/2,\bm{R})H_n(t,\,\bm{R})\phi(t,\,\bm{R},\bm{R}+\bm{r}/2)\psi^\dagger(t,\,\bm{R}+\bm{r}/2)\,,\label{intop}
\end{align}
with $H_n(\bm{R})$ a gluonic operator or light-quark operator from table~\ref{gop}, $\psi$ the Pauli spinor field that annihilates a quark, $\chi$ the one that creates an antiquark and $\phi$ is a Wilson line.

\begin{figure}[h]
\centering
\includegraphics[width=0.55\linewidth]{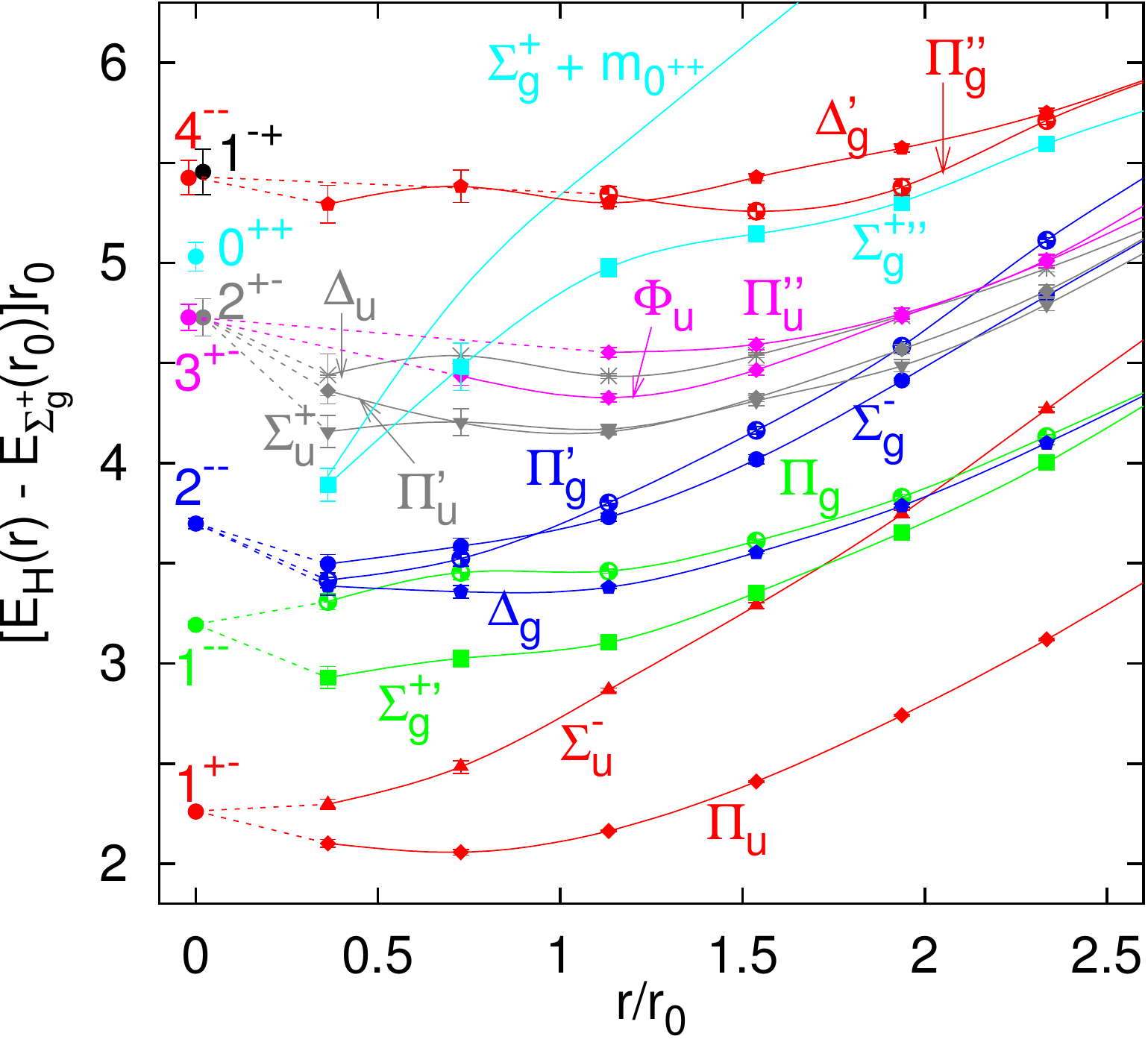}
\caption{The lowest hybrid static energies~\cite{Juge:2002br} and gluelump masses~\cite{Foster:1998wu} in units of $r_0\approx 0.5$~fm. The absolute values have been fixed such that the ground state $\Sigma_g^+$ static energy (not displayed) is zero at $r_0$.}
\label{stelat}      
\end{figure}

The correlator in Eq.~\eqref{ste} with the interpolating operator of Eq.~\eqref{intop} corresponds to a static Wilson loops with the insertion in the spatial sides of the $H_n$ light degree of freedom operator.
\begin{table}
\centering
\caption{Examples of gluonic operators and light-quark operators for quarkonium hybrids and tetraquarks respectively, $\bm{q}=(u,\,d)$ and $\tau^a$ are isospin Pauli matrices.}
\label{gop}
\begin{tabular}{ll|l|l}
$\Lambda_\eta^\sigma$   & $\kappa$ & $H$ & $H=H^aT^a(I=0,\,I=1)$ \\ \hline
$\Sigma_g^+$            & $0^{++}$ & $\mathbbm{1}$ & $\bm{\bar{q}}T^a(\mathbbm{1},\,\bm{\tau})\bm{q}$ \\
$\Sigma_u^-$            & $1^{+-}$ & ${\bf \hat{r}}\cdot{\bf B} $ & $\bm{\bar{q}}\,\left[(\hat{\bm{r}}\times\bm{\gamma})\cdot,\,\bm{\gamma}\right] T^a(\mathbbm{1},\,\bm{\tau})\bm{q}$\\ 
$\Pi_u$                 & $1^{+-}$ & ${\bf \hat{r}}\times{\bf B}$ & $\bm{\bar{q}}\,\left[\hat{\bm{r}}\cdot\bm{\gamma},\,\bm{\gamma}\right] T^a(\mathbbm{1},\,\bm{\tau})\bm{q}$\\
$\Sigma_g^{+\, \prime}$ & $1^{--}$ & ${\bf \hat{r}}\cdot{\bf E} $ & $\bm{\bar{q}}\, (\bm{\hat{r}}\cdot\bm{\gamma}) T^a(\mathbbm{1},\,\bm{\tau})\bm{q}$\\
$\Pi_g$                 & $1^{--}$ & ${\bf \hat{r}}\times{\bf E} $ & $\bm{\bar{q}}\, (\bm{\hat{r}}\times\bm{\gamma})T^a(\mathbbm{1},\,\bm{\tau})\bm{q}$\\
\end{tabular}
\end{table}
It also involves nonperturvative dynamics but it is convenient quantity to compute on the lattice. The most recent lattice results for hybrid quarkonium static energies have been computed in Refs~\cite{Juge:1997nc,Juge:2002br,Bali:2000vr,Capitani:2018rox}. In figure~\ref{stelat} we show the spectrum corresponding to the operators of table~\ref{gop}. Analogous studies on the lattice of tetraquark static energies have not been yet performed.
One important property of the static energies shown in figure~\ref{stelat}, is that they form quasi-degenerate multiplets in the short-distance region~\cite{Brambilla:1999xj,Berwein:2015vca}. This can be easily understood as an enlargement of the symmetry from a cylindrical group, $D_{\infty h}$, to an spherical one, $O(3)\times C$, in the $r\to 0$ limit. The degenerate multiplets of gluonic static energies can also be read from table~\ref{gop} corresponding to the static energies interpolated by the same gluonic operator with different projections on the heay-quark-antiquark axis. The short distance limit $O(3)\times C$ representation of the static energies can be read from the second column of table~\ref{gop}.

The results in figure~\ref{stelat} are obtained in the quenched approximation. Results with dynamical quarks are available only for first excited hybrid static energy $\Pi_u$ from Ref.~\cite{Bali:2000vr} where no statistically significant differences were found, when comparing to an equivalent computation in pure $SU(3)$ gauge theory. We expect this behavior to hold for the rest of the hybrid static energies.

It is important to note that interpolating operators of the type of Eq.~\eqref{intop} do not cover the whole spectrum of static energies as, for example, the ones associated to heavy meson pair states.
Heavy meson pairs from the first excited state for $I=0$. The heavy-meson thresholds appear in the static energy spectrum as roughly horizontal lines. The crossing of the lowest heavy-meson threshold with the ground state (standard quarkonium) static energy was studied in Ref.~\cite{Bali:2005fu} where it was shown that the effect on the static energies was tiny, apart from avoiding level crossing. Similar avoided crossings are expected for higher energy heavy meson thresholds and hybrid static energies. Another type of static state to consider are the combination of quarkonia with light-quark hadron states, these will form the ground state for $I\neq 0$. These static energies can be obtained from interpolating operators of the type of Eq.~\eqref{intop} with the color singlet analog of the operators rightmost column of table~\ref{gop}. We expect these static energies to be well approximated by the standard quarkonium static energy shifted up by the mass of the light-quark hadron~\cite{Bali:2005fu,Alberti:2016dru}. The deviation from this expectation can be interpreted as the interaction of the quarkonia with the light-quark hadron. The hadrocharmonium proposal of Ref.~\cite{Dubynskiy:2008mq} can be understood as bound states on these static energies.

Now, let us build an EFT describing heavy-quark bound states on the hybrid and tetraquark static energies. These bound states are small energy fluctuations around the minima of the static energies, thus the binding energy $E_b\sim\sqrt{\Lambda^3_{\rm QCD}/m}\ll \Lambda_{\rm QCD}$. Since $E_b\sim mv^2$, the relative distance scales as $r\sim 1/(mv)\lesssim 1/\Lambda_{\rm QCD}$. Therefore, the hybrid and tetraquark EFT is obtained from NRQCD by integrating out the $\Lambda_{\rm QCD}$ modes. For simplicity and brevity in the following we are going to consider only
the cases of $I=0$, $I=1$, and $\kappa=0^{+-},1^{+-}$ and $1^{--}$. In the $I=0$ sector the Lagrangian reads
\begin{align}
L^{(I=0)}_{BO} &= \int d^3Rd^3r \,\Bigl[S^{\dag}\left(i\partial_t-V_{\Sigma^+_g}(r)+\frac{\bnabla^2_r}{m}\right)S \nn \\
&+\sum_{\kappa=1^{+-},\,1^{--}} \sum_{\la\lap}\Psi^{\dagger}_{\kappa\lambda} \biggl\{i\partial_t - V_{\kappa\la\lap}(r)+\hat{\bm{r}}^{i\dag}_{\la}\frac{\bnabla^2_r}{m}\hat{\bm{r}}^i_{\lap}\biggr\}\Psi_{\kappa\lap}\,. \label{liso0}
\end{align}
The fields $\S$ and $\Psi_{\kappa\lambda}$ have to be understood as depending on $t$, $\bm{r}$ and $\bm{R}$. The projectors are $\hat{r}^i_0=\hat{r}^i$ and $\hat{r}^i_\pm=\mp\left(\hat{\theta}^i\pm i\hat{\phi}^i\right)/\sqrt{2}$ where $\hat{\bm r} = (\sin\theta\cos\phi,\,\sin\theta\sin\phi\,,\cos\theta)$, $\hat{\bm \theta} =$ $(\cos\theta\cos\phi,$ $\,\cos\theta\sin\phi\,,-\sin\theta)$ 
and $\hat{\bm \phi} = (-\sin\phi,\,\cos\phi\,,0)$. The first line in Eq.~\eqref{liso0} corresponds to the strongly coupled pNRQCD Lagrangian descriving the standard quarkonium spectrum\footnote{The EFT for hybrids and tetarquarks is in fact an extension of strongly coupled pNRQCD.} and the second line describes the quarkonium hybrid spectrum associated to the lowest energy hybrids static energies. The hybrid quarkonium spectrum was computed in this EFT framework in Refs~\cite{Berwein:2015vca,Soto:2017one}.

Let us define the isovector field as
\begin{align}
Z_{\kappa}=Z_{\kappa}^a\tau^a=\left(
\begin{array}{cc}
Z_{\kappa}^0 & \sqrt{2} Z_{\kappa}^+\\
\sqrt{2} Z_{\kappa}^- & -Z_{\kappa}^0 \\
\end{array}
\right)\,,
\end{align}
where $\tau^a$ are the isospin Pauli matrices. The $I=1$ sector Lagrangian is
\begin{align}
L^{(I=1)}_{BO} &= \int d^3Rd^3r \, \Bigl[\langle Z^{\dag}_{0^{+-}}\left(iD_t-V_{\Sigma^+_g}(r)+\frac{\bnabla^2_r}{m}\right)Z_{0^{+-}}\rangle \nn \\
&\sum_{\kappa=1^{+-},\,1^{--}} \sum_{\la\lap}\langle Z^{\dag}_{\kappa\lambda} \biggl\{iD_t - V_{\kappa\la\lap}(r)+\hat{\bm{r}}^{i\dag}_{\la}\frac{\bnabla^2_r}{m}\hat{\bm{r}}^i_{\lap}\biggr\}Z_{\kappa\lap}\rangle\,,\label{liso1}
\end{align}
which generates a spectra of tetraquark states. The fields $\bm{Z}$ field is understood as depending on $t$, $\bm{r}$ and $\bm{R}$. We use the notation $\langle \rangle$ to denote the trace over isospin indices. The covariant derivative for the $I=1$ fields read $D_{\mu}\bm{Z}=\pa_{\mu}+[\Gamma_{\mu},\,\bm{Z}]$ with $\Gamma_{\mu}=\left(u^{\dag}\partial_{\mu}u+u\partial_{\mu}u^{\dag}\right)/2$ and $u={\rm exp}(i\bm{\pi}\cdot\bm{\tau}/(2F))$. The pion fields depend on $t$ and $\bm{R}$.  

\subsection{Matching and short distance regime} \label{sdm}

It is interesting to study the short-distance regime in which the inter-quark distance can be considered $r\ll 1/\Lambda_{\rm QCD}$. In this case the scale associated to the relative heavy-quark momentum $mv\sim 1/r\gg \Lambda_{\rm QCD}$ can be integrated out perturbatively leading to an EFT formally identical to weakly-coupled pNRQCD~\cite{Pineda:1997bj,Brambilla:1999xf}. One can then in turn integrate out the $\Lambda_{\rm QCD}$ modes and match weakly-coupled pNRQCD to the EFT for hybrids and tetraquarks~\cite{Berwein:2015vca, Brambilla:2017uyf}. This procedure yields a short distance description of the EFT potentials. 

The matching conditions from NRQCD to weaky-coupled pNRQCD to the hybrid and tetraquark EFT are
\begin{align}
\mathcal{O}_n(t,\,\bm{r},\,\bm{R})\cong Z_{H_n}(r) O^a(t,\,\bm{r},\,\bm{R})H^a_n(t,\,\bm{R})+\dots\cong Z_{\mathcal{X}_n}(r,\,\Lambda_{\rm QCD})\mathcal{X}_{n}(t,\,\bm{r},\,\bm{R})+\dots
\end{align} 
where $\mathcal{X}_n$ stands for $\Psi_n$ or $Z_n$, $O^a$ is the color-octet heavy quark-antiquark field. $Z_{H_n}$ and $Z_{\mathcal{X}_n}$ are normalization factors.

Generically, in the short distance the potentials may be organized as a sum of a perturbative part, which is typically nonanalytic in $r$ corresponding to the weakly-coupled pNRQCD potentials, and a nonperturbative part, which is a series in powers of $r$. The coefficients of the latter only depend on $\Lambda_{\rm QCD}$ and can be expressed in terms of gluonic and light-quark correlators. The static potential, $V^{(0)}_{\kappa\la}$, can be matched to the lattice NRQCD static energies and a short distance weak-coupling pNRQCD description:
\begin{align}
E^{(0)}_{|\lambda|^{\sigma}_{CP}}(r)=V_o(r)+\Lambda_{\kappa}+b_{\kappa\lambda}r^2+\dots=V^{(0)}_{\kappa\la}(r)\,,
\end{align} 
where $V_o(r)$ is the octet potential, $\Lambda_{\kappa}$ is the gluelump mass~\cite{Foster:1998wu}, and $b_{\kappa\lambda}$ is a nonperturbative constant.

\begin{figure}[h]
\centering
\includegraphics[width=1\linewidth]{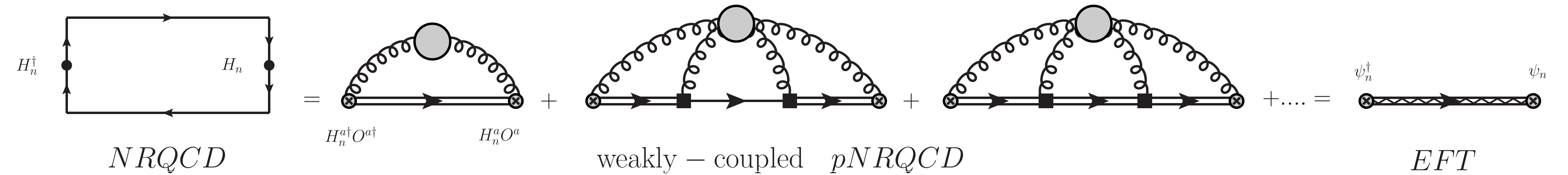}
\caption{Matching diagram for the static potential. The double, single and curly lines represent the heavy-quark-antiquark octet, singlet and the gluon field respectively. The black squares stands for multipolar vertices from the weakly coupled pNRQCD Lagrangian, the circles with a cross for the light degree of freedom operators and the shaded circle represents nonperturbative gluon exchanges.}
\label{sp_matching}       
\end{figure}

\subsection{Spin-dependent terms for quarkonium hybrids}
The potential $V_{\kappa\lambda\lambda^{\prime}}$ can be organized into an expansion in $1/m$ and a sum of spin-dependent ({\rm SD}) and spin-independent parts~\cite{Brambilla:2018pyn,Soto:2017one}:
\begin{align}
V_{\kappa\la\lap}(\bm{r}) = V^{(0)}_{\kappa\la}(r)\de_{\la\lap}+\frac{V^{(1)}_{\kappa\la\lap}(\bm{r})}{m}+\frac{V^{(2)}_{\kappa\la\lap}(\bm{r})}{m^2}+\dots
\end{align}
For the lowest-lying hybrid excitations of $\kappa=1^{+-}$, the spin-dependent potentials take the form 
\begin{align}
V_{1^{+-}\la\lap\,{\rm SD}}^{(1)}(\bm{r}) =& V_{{\rm SK}}(r)\left(\hat{r}^{i\dag}_{\la}\bm{K}^{ij}\hat{r}^j_{1\lap}\right)\cdot\bm{S}+ V_{{\rm SK}b}(r)\left[\left(\bm{r}\cdot \hat{r}^{\dag}_{\la}\right)\left(r^i\bm{K}^{ij}\hat{r}^j_{\lap}\right)\cdot\bm{S}\right.\nn\\
&\left.+\left(r^i\bm{K}^{ij}\hat{r}^{j\dag }_{\la}\right)\cdot\bm{S} \left(\bm{r}\cdot \hat{r}_{\lap}\right)\right]+\dots \label{sdm2}\\
V_{1^{+-}\la\lap\,{\rm SD}}^{(2)}(\bm{r}) =& V_{{\rm SL}a}(r)\left(\hat{r}^{i\dag}_{\la}\bm{L}_{Q\bar{Q}}\hat{r}^i_{\lap}\right)\cdot\bm{S}\nn+ V_{{\rm SL}b}(r)\hat{r}^{i\dag}_{\la}\left(L_{Q\bar{Q}}^iS^j+S^iL_{Q\bar{Q}}^j\right)\hat{r}^{j}_{\lap}+ V_{{\rm S}^2}(r)\nn\\
&\bm{S}^2\de_{\la\lap}+V_{{\rm S}_{12}a}(r)S_{12}\de_{\la\lap}+ V_{{\rm S}_{12}b}(r)\hat{r}^{i\dag}_{\la}\hat{r}^j_{\lap}\left(S^i_1S^j_2+S^i_2S^j_1\right)+\dots\label{sdm3}
\end{align}
where $\bm{L}_{Q\bar{Q}}$ is the orbital angular momentum of the heavy-quark-antiquark pair, $\bm{S}_1$ and $\bm{S}_2$ are the spin vectors of the heavy quark and heavy antiquark, respectively, $\bm{S}=\bm{S}_1+\bm{S}_2$ and ${S}_{12}=12(\bm{S}_1\cdot\hat{\bm{r}})(\bm{S}_2\cdot\hat{\bm{r}})-4\bm{S}_1\cdot\bm{S}_2$. $\left({K}^{ij}\right)^k=i\epsilon^{ikj}$ is the angular momentum operator for the spin-1 representation.  Standard time-independent perturbation theory can be used to compute the mass shifts in the quarkonium hybrid spectrum produced by the spin-dependent operators.

\begin{figure}[h]
\centering
\includegraphics[width=0.42\linewidth]{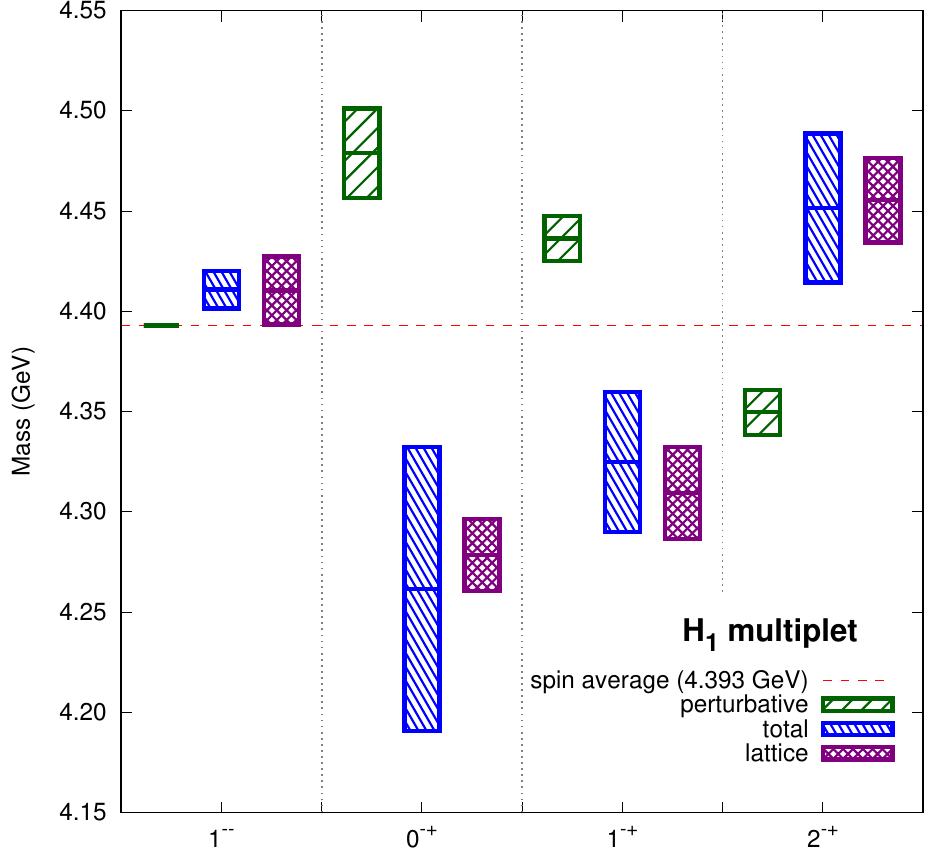}
\includegraphics[width=0.42\linewidth]{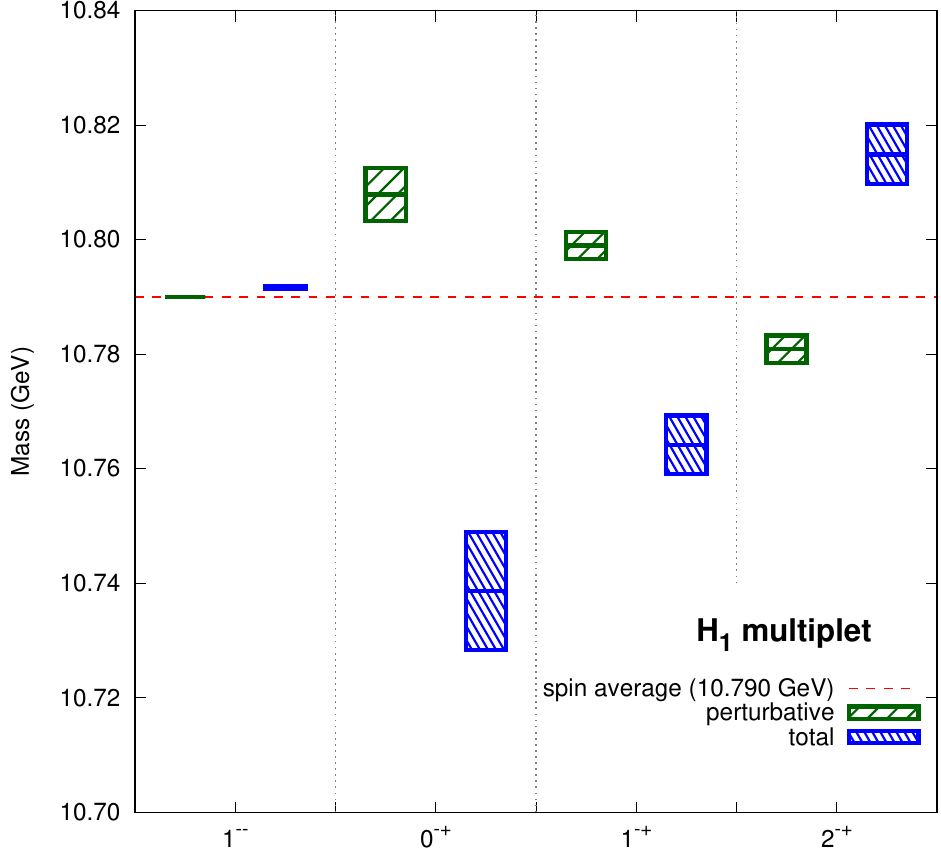}
\caption{Spectrum of the lowest-lying charmonium (left) and bottomonium (right) hybrid multiplet. The lattice results for the charmonium sector from Ref.~\cite{Cheung:2016bym} are the most right (purple) boxes for each quantum number. The perturbative contributions to the spin-dependent operators in Eq.~\eqref{sdm3} added to the spin average of the lattice results (red dashed lines) are the most left (green) boxes. The central (blue) boxes for each quantum number are the full results from the spin-dependent operators of Eqs.~\eqref{sdm2} and \eqref{sdm3} including perturbative and nonperturbative contributions. The height of the boxes indicates the uncertainty.}
\label{spspl}     
\end{figure}

In the short-distance approximation the matching coefficients of the spin-dependent potentials can be determined up to a set of nonperturbative constants. These constants can be obtained by fitting the spin splittings to the lattice determinations of the charmonium hybrid spectrum. Once the nonperturvative part of the matching coefficients are determined, the spin contributions in the bottomonium hybrid sector can be predicted~\cite{Brambilla:2018pyn}. In figure~\ref{spspl} we show the results for the lowest-lying hybrid multiplet.

\subsection{Mixing with standard Quarkonia}

A reliable study of the exotic quarkonium spectrum and its properties should take into account the mixing between static states that exist in the same energy range~\cite{Oncala:2017hop}. The Lagrangians in Eq.~\eqref{liso0} and \eqref{liso1} incorporate the mixing between static states with degenerate short-distance limit but not with other states in different short-distance $\kappa$ representations neither different types of static states such as the ones corresponding to heavy meson pairs. The mixing operators for $\kappa=1^{+-},\,1^{--}$ hybrids and quarkonia read
\begin{align}
L^{\rm mixing}_{BO} =& \int d^3Rd^3r \,\Bigl[M_{11}(r)\left(S^{\dag}\left\{\bm{\sigma}\cdot\hat{\bm{r}}_{\la},\Psi_{1^{+-}\la}\right\}+\text{h.c}\right)+M_{12}(r)\left(S^{\dag}\left\{(\bm{\sigma}\cdot r)(r\cdot\hat{\bm{r}}_{\la}),\right.\right.\nn\\
&\left.\left.\Psi_{1^{+-}\la}\right\} +\text{h.c}\right)+M_{21}\left(\bm{r}\cdot\hat{\bm{r}}_{\la}S^{\dag}\Psi_{1^{--}\la}+\text{h.c}\right)\Bigr]\,.
\end{align}
So far the mixing only the mixing of the lowest lying hybrid ($\kappa=1^{+-}$) and quarkonium static energies has been studied in Ref.~\cite{Oncala:2017hop}. The mixing potential was studied in the short-distance regime using the intermediate step of matching to weaky-coupled pNRQCD as detailed in Sec.~\ref{sdm}, and in the long-distance regime ($r\gg 1/\Lambda_{\rm QCD}$) by evaluating the Wilson loops involved in the matching using an Effective String Theory. The interpolation between both regimes was modeled. The quarkonium component, up to $30\%$, resulted to be fairly significant in some states. The mixing term implies that the actual physical states are a superposition of spin zero (one) hybrids and spin one (zero) quarkonium. This facilitates the identification of certain exotic states as hybrids, since otherwise the apparent spin symmetry violating decays were difficult to understand. 

\subsection{Semi-inclusive hybrid decays}

If the energy gap $\Delta E$ between a hybrid and a low-lying standard quarkonium state is large $\Delta E\geq \Lambda_{\rm QCD}$ it is possible to obtain a perturbative estimate of the semi-inclusive transition. The low-lying standard quarkonium states can be integrated out generating an imaginary potential in the hybrid EFT which in turn will produce the semi-inclusive decay width for a hybrid state to decay into a given low-lying quarkonium state. In this setting, the imaginary potential corresponds to the imaginary part of the octet self energy in weakly coupled pNRQCD~\cite{Oncala:2017hop}.
\begin{align}
\Gamma(\Psi_{m}\to S_n)=\frac{4}{3}\frac{\alpha_s T_F}{N_c}|\langle \Psi_{m}|\bm{r}| S_n \rangle|^2(\Delta E_{mn})^3\,.
\end{align}
The matrix element selects hybrid states with $\Sigma$ component, $\Delta l=0$, and $\Delta S=0$. Numerical evaluations of the semi-inclusive width can be found in Ref.~\cite{Oncala:2017hop}.

{\bf Acknowledgments} \\ 
This work was supported by the Spanish grants FPA2017-86989-P and SEV-2016-0588 and by the European Union's Horizon 2020 research and innovation programme under the Marie Sk\l{}odowska--Curie Grant Agreement No. 665919.

\bibliography{bibcharm18}
\end{document}